\documentclass[twocolumn,aps,prl,showpacs,epsfig]{revtex4}
\usepackage{graphicx}

\begin{document}

\title{Critical Current of the Spin-Triplet Superconducting Phase in Sr$_2$RuO$_4$}

\author{Hae-Young Kee$^1$}
\author{Yong Baek Kim$^1$}
\author{K. Maki$^2$}
\affiliation{${}^1$Department of Physics, University of Toronto, Toronto, Ontario
M5S 1A7, Canada\\
${}^2$Department of Physics, University of Southern California, Los Angeles,
CA 90089}
\date{\today}

\begin{abstract}
There have been two different proposals for the spin-triplet order 
parameter of the superconducting phase in Sr$_2$RuO$_4$; 
an $f$-wave order parameter
and the multigap model where some of the bands have the line node. 
In an effort to propose an experiment that can distinguish two
cases, we study the behavior of the supercurrent and compute 
the critical current for these order parameters when the
sample is a thin film with the thickness $d \ll \xi$ where
$\xi$ is the coherence length. It is found that
the supercurrent behaves very differently in two models.
This will serve as a sharp 
test for the identification of the correct order parameter.
\end{abstract}


\maketitle

{\bf Introduction}: The order parameter of the superconducting phase 
in Sr$_2$RuO$_4$ has been a subject of intensive research since its 
discovery in 1994 \cite{maeno,mackenzie}. 
In particular, the field has been stimulated by 
the prospect of identifying the simplest electronic version of
spin-triplet superconductor. The early proposal made by Rice and 
Sigrist suggested the following spin-triplet order parameter with 
$p$-wave symmetry \cite{rice,maeno2}.
\begin{equation}
{\hat \Delta}({\bf k})= \Delta {\hat d} \ (k_x \pm i k_y),
\end{equation}
where $\Delta$ is the magnitude of the order parameter and
${\hat d}$ is a unit vector perpendicular to the spin of the 
condensed pair \cite{vollhardt}. 
This order parameter breaks time-reversal 
symmetry and has a full gap on the Fermi surface.
Indeed the flat ${}^{17}$O Knight shift 
across $T_c$ \cite{ishida}, for the magnetic field 
along the $ab$ plane, and the observation of 
spontaneous magnetic moment in $\mu$SR \cite{luke}
are consistent with the spin-triplet order parameter and broken 
time-reversal symmetry, respectively.
This encouraged theoretical investigations of
the effects of the characteristic collective 
modes (the spin waves and clapping mode) and 
the topological defects with the order 
parameter of Eq.1 \cite{tewordt,kee,kee2,half}. 
 
The experiments performed later on cleaner samples, however,
reveal the existence of the nodal structure in the order
parameter; the $T^2$ dependence of the specific heat \cite{nishizaki},
the $T$-linear behavior of the superfluid density \cite{bonalde}, 
the $T^3$ dependence of $1/T_1$ in NMR \cite{ishida3}, the $T^2$ behavior
of the ultrasonic attenuation \cite{lupien}, and the $\sqrt{H}$ dependence
of the specific heat in a magnetic field \cite{nishizaki} at low 
temperatures. While these behaviors are consistent with
an $f$-wave superconductor with nodes \cite{hasegawa,won,dahm,miyake,graf}, 
this may be at odd with the fact that the
quasi-two dimensional system with strong ferromagnetic
fluctuations would typically favor the $p$-wave superconductor
with Eq.1 \cite{sigrist,sato}. In an effort to resolve this issue, the following
order parameter with the horizontal node was proposed as 
a strong candidate \cite{won}.
\begin{equation}
{\hat \Delta}({\bf k})= \Delta {\hat d} \ (k_x \pm i k_y) \
{\rm cos} (ck_z),
\end{equation}
where $c$ is the lattice constant in the $c$-axis.   
It was shown that the magneto-thermal conductivity would
provide useful information about the validity of this
order parameter \cite{won2,won3}. Such experiments were indeed 
carried out \cite{tanatar,izawa},
and the results are consistent with Eq.2, even though the 
data cannot exclude the possible presence of small amount of
$p$-wave mixture.   

An alternative model was suggested by 
Zhitomirsky and Rice \cite{zhito,annett}.
Their multigap model utilizes the fact that there exist 
three bands; $\alpha$, $\beta$, and $\gamma$ \cite{maeno2}. 
Here the dominant $p$-wave order parameter with Eq.1 resides on the 
active $\gamma$ band while the proximity effect leads to
the following $f$-wave-like (it will be called $f'$-wave 
hereafter) order parameter with the line nodes 
in the $\alpha$ and $\beta$ bands \cite{zhito}.
\begin{equation}
{\hat \Delta}({\bf k})= \Delta {\hat d} \ (k_x \pm i k_y) \
{\rm cos} (ck_z/2).
\end{equation}
The observation of a double transition in a recent specific 
heat measurement {\it near} $H_{c2}$ 
was interpreted in terms of 
this multigap model \cite{deguchi}. 
On the other hand, the specific heat and the magnetic
penetration depth for low temperatures ($T \ll T_c$) and
{\it low} field ($H \ll H_{c2}$) appear to be consistent with
the $f$-wave order parameter of Eq.2.

Thus it is not clear at the moment which order parameter
is the correct description of the superconducting phase; 
this is the fundamental issue for Sr$_2$RuO$_4$.
Previously it was suggested that the angle dependence of
the magneto-thermal conductivity can be used to distinguish 
different order parameters \cite{maki}. The Raman spectra can be also
used to detect different contributions from the clapping mode
in the $f$-wave and the multigap models \cite{kee3,dora}. 
These proposals, however, reply on the {\it quantitative}, 
albeit detectable, 
difference in two models. Thus it is still desirable to
have a sharp test that leads to the {\it qualitatively} 
different outcome in two cases.   

In this paper, we provide such a proposal; the study of 
the critical current in two models for the spin-triplet 
superconductor. In particular, we investigated the
behavior of the supercurrent for the $f$-wave and $f'$-wave 
order parameters. It is found that the supercurrent
in the $ab$-plane has the same form in both cases
and has significant contributions from the quasiparticles.
On the other hand, if the current flows along the
$c$-axis, the supercurrent in the $f'$-wave is relatively
immune to the quasiparticles at small current 
while that of the $f$-wave still acquires a large
contribution from the quasiparticles.  
Notice that the supercurrent in
the multigap model would be dominated by the
$p$-wave component on the $\gamma$-band 
and its behavior would be the same as
the case of $s$-wave superconductor where
the quasiparticle contribution is almost absent
at small current \cite{parks}. 
Combining all these, it can be seen 
that the supercurrent in the $f$-wave case 
is much more affected by the quasiparticles.
Thus the study of the supercurrent provides a clear 
mean to distinguish two different proposals for 
the order parameter
of the superconducting phase in Sr$_2$RuO$_4$.

More specifically, we will consider a thin film with 
the sample thickness $d \ll \xi$, where $\xi$ is the 
coherent length. Under this condition, the magnitude of
the superconducting gap, $\Delta$, and the
supercurrent will be uniform across the system \cite{parks}.
When a uniform current flows, the Cooper-pair
acquires a center of mass momentum of ${\bf q}_s$.
If there were no contribution from the quasiparticles, 
the supercurrent would be simply proportional to ${\bf q}_s$.
The quasiparticles, however, change this behavior via
the shifted quasiparticle dispersion, $E_{\bf k}$, in the 
presence of the superflow \cite{parks}.
\begin{equation}
E_{\bf k} = E^0_{\bf k} + {\bf v}_{\bf k} \cdot {\bf q}_s,
\end{equation}
where $E^0_{\bf k} = \sqrt{\xi^2_{\bf k} + \Delta^2_{\bf k}}$
and ${\bf v}_{\bf k} = \partial \xi_{\bf k} / \partial {\bf k}$
is the quasiparticle velocity. 
The total current, $j_s$, as a function of $s = v_F q_s$ rises 
linearly at small $s$ and has a maximum at some value of 
$s$, then drops as $s$ is further increased. 
When $dj_s/ds < 0$, the system is unstable;
thus the critical current is determined by the value of $s$
where $dj_s/ds = 0$ \cite{parks}.
The detailed behavior of the supercurrent and the value
of the critical current in each case are discussed below.

{\bf $f$-wave order parameter}: We will consider two different
cases; the current in the $ab$-plane and along the $c$-axis.

\noindent
{\bf a. the current in the $ab$-plane}

The gap equation for the order parameter in the 
presence of a uniform current can be written as
\begin{equation}
\Delta({\bf k}) =  - T \sum_{i \omega_n} \sum_{\bf p} 
V({\bf k},{\bf p}) {\rm Tr}[ \rho_1 \sigma_1 G (i\omega_n, {\bf p})]
\end{equation}
where $\Delta({\bf k}) = \Delta f({\bf k})$,
$V({\bf k},{\bf p}) = V f({\bf k}) f({\bf p})^*$,
and $f({\bf k})$ represents the momentum dependence of
the order parameter.
Here the single particle Green's function $G (i\omega_n, {\bf k})$ 
is given by 
\begin{equation}
G^{-1}(i\omega_n,{\bf k}) =i \omega_n -{\bf v}_{\bf k} 
\cdot {\bf q}_s +
\xi_{\bf k} \rho_3 +\Delta({\bf k}) \rho_1 \sigma_1,
\end{equation}
where $\xi_{\bf k} = (k_x^2+k_y^2)/2m - t \cos(c k_z) -\mu$
is the single particle dispersion. 
 
The uniform current reduces the amplitude of the order 
parameter and the current dependence of the amplitude at $T=0$ 
can be obtained from the following equation derived from Eq.4
(the lattice constant $c=1$ for simplicity).
\begin{equation}
\ln \left [ \frac{\Delta(0)}{\Delta(s)} \right ] =
\frac{8}{\pi^2} \int_0^{\pi \over 2} d k_z \int_0^{\pi \over 2} 
d\phi \ |f|^2 \ {\rm Re} \ 
{\rm arccosh}\left [ \frac{s \sin\phi }{\Delta(s) |f|} \right ]
\end{equation}
where $s= v_F q_s$, $\phi$ is the angle between the direction of the 
current ${\hat {\bf q}}_s$ and the quasiparticle velocity ${\bf v}_{\bf k}$.
$\Delta(s)$ and $\Delta(0)$ represent the amplitude of the order 
parameter in the presence and absence of the current, respectively.

\begin{figure}[h]
\includegraphics[height=15cm,width=7cm,angle=0]{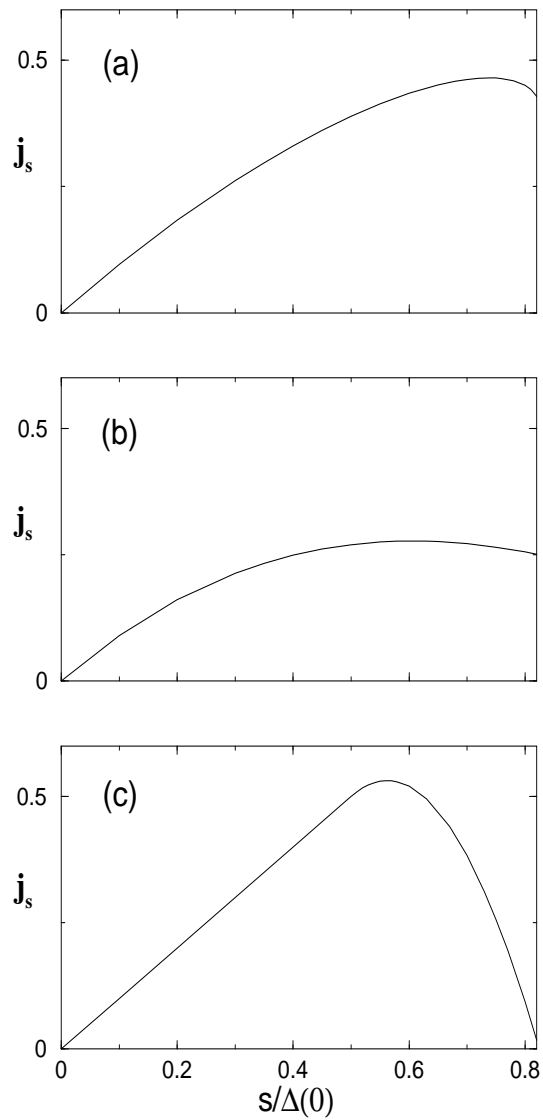}
\caption{The supercurrent as a function of $s/\Delta(0)$ with the
current (a) in the $ab$-plane for both of the $f$-wave 
and $f'$-wave, (b) along the $c$-axis for the $f$-wave, 
(c) along the $c$-axis for the $f'$-wave. The unit of the supercurrent
is $(en/mv_F)\Delta(0)$ and $(en/m)(v_{Fc}/v^2_F)\Delta(0)$ for 
the current in
the plane and along the $c$-axis, respectively. 
}
\label{lattice}
\end{figure}

In the case of the $f$-wave order parameter, 
$f({\bf k})=e^{\pm i \phi} \cos (ck_z)$, 
a straightforward computation at $T=0$ leads to
\begin{widetext}
\begin{equation}
\ln \left [ \frac{\Delta (0)}{\Delta (s)} \right ] =
\frac{2}{\pi}
\left [ \arcsin y \left ( \ln y-\frac{1}{2} \right )
-\frac{y}{2} \sqrt{1-y^2}  
-\int_0^{\arcsin y } d\phi \ \ln(\sin\phi) \right ]
\label{gap1}
\end{equation}
for  $y \equiv s/\Delta(s) < 1$. When $y > 1$, there is
no solution for the gap equation.
Now the contribution from the quasiparticles to the 
current can be computed from \cite{parks}
\begin{equation}
{\bf j}_{qp} =
e T \sum_{i\omega_n} \sum_{\bf k}   
{\rm Tr} [ {\bf v}_{\bf k} G(i \omega_n, {\bf k})] .
\end{equation} 
Taking into account this, the net 
supercurrent for
the $f$-wave order parameter is obtained as
\begin{eqnarray}
{\bf j}_s &=&
\frac{en}{m}{\bf q}_s \left [
1- \frac{8}{\pi^2 y} 
\int_0^{\pi \over 2} d k_z 
\int_0^{\pi\over 2} d\phi \ \cos\phi \ 
{\rm Re} \sqrt{(y \cos\phi)^2 - (\cos k_z)^2} \right ] \cr
&=& \frac{en}{m} {\bf q}_s
\left [ 1- \frac{1}{\pi y} 
 \left ( \sqrt{1-y^2}  - \frac{1-2y^2}{y} \arcsin y  
\right ) \right ], 
\label{current1}
\end{eqnarray}
where $n = k_F^2/2\pi$ is the density of electrons in the plane.
The supercurrent as a function of $s/\Delta(0)$ can be obtained from
Eq.\ref{gap1} and Eq.\ref{current1}; the result is plotted in Fig. 1(a). 
The critical current occurs at $s = 0.74 \Delta(0)$ and
the value of the critical current is $j_c = 0.465 (en/mv_F)\Delta(0)$. 

\noindent
{\bf b. the current along the c-axis}

When the current is along the $c$-axis, 
the gap equation for the $f$-wave order parameter 
at $T=0$ is now given by
\begin{equation}
\ln \left [ \frac{\Delta(0)}{\Delta(s)} \right ] 
= \frac{4}{\pi} \int_0^{\pi \over 2} d k_z \ (\cos k_z)^2 \ {\rm Re} \ 
{\rm arccosh} \left ( \frac{s \sin k_z}{\Delta \cos k_z} \right ) 
= {\rm arcsinh} \ y - \frac{y}{\sqrt{1+y^2}},
\label{gap2}
\end{equation}
where $s = v_{Fc} q_s$ with $v_{Fc} = tc$ and 
${\bf v}_{\bf k}= (tc) \sin k_z {\hat {\bf q}}_s$ is used.
The net supercurrent is also found as
\begin{equation}
{\bf j}_s =
\frac{en}{m} \left ( \frac{v_{Fc}}{v_F} \right )^2 {\bf q}_s 
\left [ 1 - \frac{4}{\pi y} \int_0^{\pi \over 2} 
d k_z \sin k_z \ {\rm Re}  
\sqrt{ (y \sin k_z)^2 - (\cos k_z)^2} \right ] 
= \frac{en}{m} \left ( \frac{v_{Fc}}{v_F} \right )^2 
{\bf q}_s \left [ 1 - \frac{y}{\sqrt{1+ y^2}} \right ].
\label{current2}
\end{equation}
Notice that $v_{Fc}$ and $v_F$ are the velocities along the 
$c$-direction and in the $ab$-plane, respectively.
The supercurrent computed from Eq.\ref{gap2} and Eq.\ref{current2}
is plotted in Fig. 1(b); the critical current occurs at 
$s = 0.61 \Delta(0)$ and the value of the critical
current is $j_c = 0.277 (en/m) (v_{Fc}/{v_F}^2) \Delta(0)$.

{\bf $f'$-wave order parameter}:
When the current is in the $ab$-plane, the gap equation and the expression for
the supercurrent turns out to be the same as those of the $f$-wave case.
Thus the supercurrent in this case is given by Fig. 1(a). 
On the other hand, when the current is along the $c$-axis, both of
the gap equation and the supercurrent for the $f'$-wave order parameter
are quite different from the $f$-wave counterpart. The gap equation 
now has the following form with $s=v_{Fc}q_s$.
\begin{eqnarray}
\ln \left [ \frac{\Delta(0)}{\Delta(s)} \right ] &=&
\frac{4}{\pi} \int_0^{\pi \over 2} d k_z 
\left [ \cos \left ( {k_z \over 2} \right ) \right ]^2 {\rm Re} \  
{\rm arccosh} \left [ \frac{s \sin k_z}{\Delta \cos ({k_z \over 2})} \right ] \cr 
&=& \left [ \ln (2 y) -\frac{1}{2} \left ( 1-\frac{1}{4 y^2} \right )
\right ] \theta \left ( y-\frac{1}{2} \right ),
\label{gap3}
\end{eqnarray}
where $\theta(x) = 1$ for $x > 0$ and is zero if $x < 0$.  
The supercurrent is given by
\begin{eqnarray}
{\bf j}_s &=&
\frac{en}{m} \left ( \frac{v_{Fc}}{v_F} \right )^2 {\bf q}_s  
\left [ 1- \frac{2}{\pi y} \int_0^{\pi}  d k_z
 \sin k_z \ {\rm Re} \sqrt{ (y \sin k_z)^2-\left (\cos {k_z \over 2} 
\right )^2 } \right ] \cr
&=& \frac{en}{m} \left ( \frac{v_{Fz}}{v_F} \right )^2 {\bf q}_s  
\left [ 1- \left ( 1- \frac{1}{4y^2} \right )^2 
\theta \left ( y-\frac{1}{2} \right )\right ] .
\end{eqnarray}
\noindent
The supercurrent as a function of $s/\Delta(0)$ is plotted in Fig. 1(c).
Notice that the quasiparticle contribution to the current does not
enter for $s/\Delta < 1/2$, so the supercurrent is
proportional to ${\bf q}_s$ in this regime.
The critical current is found at $s = 0.56 \Delta(0)$ 
and the value 
of the critical current is $j_c = 0.531 (en/m) (v_{Fc}/{v_F}^2) \Delta(0)$.

\vskip 0.3cm

\end{widetext}

{\bf Summary and Conclusion}:
We investigated the behavior of the supercurrent and 
obtained the critical current in the cases of the 
$f$-wave and $f'$-wave order parameters defined in
Eq.2 and Eq.3. In the superconductor with the full gap, the 
quasiparticle contribution to the supercurrent would
enter only when the current exceeds some value.
If the order parameter has nodes, however, the 
quasiparticle contribution may affect the supercurrent
even in the small current limit.
When the current flows in the $ab$-plane, the 
supercurrent in the $f$-wave and $f'$-wave cases
is indeed affected by quasiparticles
even for small current and it has the same form in both cases.
On the other hand, if the current is along the $c$-axis, 
the supercurrent in the $f'$-wave case behaves similarly 
to the case with the full gap; in contrast, the quasiparticle
contribution still enters at small current in 
the $f$-wave case. 
 
Now the discussion about different models for the 
superconducting order parameter in Sr$_2$RuO$_4$ is in order.

1) When the current is in the $ab$-plane, 
the supercurrent in the multigap model would be mainly
determined by the dominant $p$-wave component on the
$\gamma$-band and its behavior is similar to the
case of $s$-wave superconductor; the small contribution from
the $f'$-wave component is subdominant so that the supercurrent
is not much affected by the quasiparticles at small current.
Thus, the critical current in the multigap model
would be much bigger than that of the $f$-wave model 
where the supercurrent acquires a significant quasiparticle 
contribution. 

2) If the current
is along the $c$-axis, even the supercurrent from the
$f'$-wave component (as well as the $p$-wave component) 
in the multigap model does not acquire the quasiparticle
contribution at small current and as a result the supercurrent 
is even less affected by the quasiparticles. On the other hand, 
the supercurrent in the $f$-wave model is still very much 
affected so that the critical current is again much smaller.
 
The qualitative difference in the 
behavior of the supercurrent in two models can be used to 
discriminate one of the leading candidates for the order 
parameter of the superconducting phase in Sr$_2$RuO$_4$.

{\it Acknowledgments}:
HYK and YBK thank Aspen Center for Physics for its hospitality
during the summer workshop in 2003.
This work was supported by the Natural Sciences and Engineering 
Research Council of Canada, Canadian Institute for
Advanced Research, Canada Research Chair Program, and the Alfred P. Sloan
Fellowship (HYK and YBK).

\end{document}